\documentclass[aps,preprint,epsf,epsfig]{revtex4}
\usepackage{graphicx,graphics,epsf,psfig,epsfig}
\usepackage{subfigure}
\def\beq{\begin{equation}}
\def\eeq{\end{equation}}
\def\beqa{\begin{eqnarray}}
\def\eeqa{\end{eqnarray}}
\begin{document}
\title{Geometry of proteins: Hydrogen bonding, sterics and marginally
compact tubes}

\author{Jayanth R. Banavar}
\affiliation{Department of Physics, 104 Davey Lab,
The Pennsylvania State University, University Park PA 16802, USA}

\author{Marek Cieplak}
\affiliation{Institute of Physics, Polish Academy of Sciences, 02-668
Warsaw, Poland}

\author{Alessandro Flammini}
\affiliation{School of Informatics, Indiana University,
Bloomington, IN 47408, USA}

\author{Trinh X. Hoang}
\affiliation{Institute of Physics and Electronics,
VAST, 10 Dao Tan, Hanoi, Vietnam}

\author{Randall D. Kamien}
\affiliation{Department of Physics and Astronomy,
University of Pennsylvania, Philadelphia, PA 19104, USA}

\author{Timothy Lezon}
\affiliation{Department of Physics, 104 Davey Laboratory,
The Pennsylvania State University, University Park PA 16802, USA}

\author{Davide Marenduzzo}
\affiliation{Mathematics Institute, University of Warwick,
Coventry CV4 7AL, England}

\author{Amos Maritan}
\affiliation{Dipartimento di Fisica `G. Galilei',
Universit\`a di Padova, Via Marzolo 8, 35131 Padova, Italy\\
Sezione INFN, Universit\`a di Padova, I-35131 Padova, Italy}

\author{Flavio Seno}
\affiliation{Dipartimento di Fisica `G. Galilei',
Universit\`a di Padova, Via Marzolo 8, 35131 Padova, Italy\\
Sezione INFN, Universit\`a di Padova, I-35131 Padova, Italy}

\author{Yehuda Snir}
\affiliation{Department of Physics \& Astronomy,
University of Pennsylvania, Philadelphia, PA 19104, USA}

\author{Antonio Trovato}
\affiliation{Dipartimento di Fisica ``G. Galilei''\\
Sezione INFN, Universit\`a di Padova, I-35131 Padova, Italy}

\begin{abstract}

The functionality of proteins is governed by their structure in the
native state. Protein structures are made up of emergent building
blocks of helices and almost planar sheets. A simple coarse-grained
geometrical model of a flexible tube barely subject to compaction
provides a unified framework for understanding the common character
of globular proteins. We argue that a recent critique of the tube
idea is not well founded.

\end{abstract}

\maketitle

\newcounter{ctr}
\setcounter{ctr}{1}

The protein problem\cite{Creighton,Fershtbook,Finkelbook} is one of
formidable complexity. The number of degrees of freedom of the protein
atoms as well as the surrounding water molecules, which play an
essential role in the folding process, is enormous. In addition, a
protein chain is relatively short compared to macromolecular polymer
chains and one might therefore expect significant non-universal
behavior with the details mattering a great deal. Furthermore, the
sequences of proteins have been subject to evolution and natural
selection, a history dependent process.  Yet there are striking
patterns that one observes in protein behavior.

All proteins fold rapidly and reproducibly\cite{Anfinsen} and their
native state structures are made of common building blocks: helices
and zig-zag strands assembled into almost planar sheets. For globular
proteins to serve vital enzymatic roles, their folded structures need
to be flexible.  The total number of distinct folds adopted by
globular proteins is only of the order of a few
thousand\cite{Chothia1}, a remarkably small number compared to the
profusion of structures one might have expected for compact chains
comprising a few hundred monomers. Furthermore, it is believed that
the folds are evolutionarily conserved\cite{Denton,Chothia2}. Many
protein sequences adopt the same native state
conformation\cite{Manytoone}. Once a sequence has selected its native
state structure, it is able to tolerate a significant degree of
mutability except at certain key locations\cite{Design}.

It has been suggested that these common attributes of globular
proteins\cite{Bernal,BanavarRMP,HoangPNAS,PRE} reflect a deeper
underlying unity in their behavior. Yet, a protein molecule along with
the surrounding water molecules constitutes a system of great
complexity. Such a system can be described at many levels. At the
finest level, one would simply treat the entire system with all the
degrees of freedom with the laws of quantum mechanics. The
difficulties associated with a first-principles quantum mechanical
approach include the large number of degrees of freedom; the necessity
of calculating the interactions during the dynamical process of
folding, with the solvent taken into account in an accurate manner;
and, even if the interactions were known exactly, the limitations of
present-day computers in being able to accurately follow the dynamics
through the folding process. Understanding such a system at this level
of description is a daunting task and has not yet been achieved.

Any alternative coarse-graining procedure implies the determination of
effective interactions that are postulated to arise on integrating out
the degrees of freedom of the water. For example, Pitard {\it et
al.}\/\cite{Pitard} have studied the folding and anisotropic collapse
of a microscopic continuous model of a homopolymer chain where each
monomer carries a dipole moment. In an equilibrium description of any
such coarse-grained model, the effective potential not only depends on
the protein conformation as represented by the values of the
coordinates of the atoms of the protein but is also a function of the
temperature. The averaging is envisioned to be carried out under the
assumption of an instantaneous equilibration of the fine details
represented by the coordinates of the water molecules. However, the
folding of a protein is not an equilibrium situation but entails
dynamical processes that cannot be captured within an equilibrium
description.

The helix is a natural, compact conformation of a short, flexible
tube. This motivated
us\cite{HoangPNAS,PRE,MaritanNature,BMMT02,CPU,Thick,Lezon} to
investigate the phase behavior of a flexible tube subject to
compaction in order to investigate whether it is related to and can
explain protein behavior. The tube is anisotropic and may be thought
of as the continuum limit of a discrete chain of discs or coins.
Unlike a chain of spheres, a chain of coins accurately captures the
symmetry of a chain molecule because associated with each object along
the chain is a special local axis defined by the tangent to the chain
and represented by the axis perpendicular to the face of the disc. The
amino acids have side chains which stick out in a direction lying
approximately in the plane of the disc.  Unlike an ordinary garden
hose, the tube is one in which each disc orients itself in such a way
that the side chain sticks out at an angle of around $143^o$ from the
normal vector \cite{PL} joining the disc center to the center of the
circle passing through the center of the disc and the centers of its
two adjacent neighbors. The tube model does not arise from an
integration of some of the degrees of freedom of a microscopic model.

For a short discrete tube, with less than 20 residues (with the same
bond length and typical thickness of a polypeptide chain), helices and
planar hairpins and sheets are found to be the preferred structures in
a marginally compact phase in which the attractive forces promoting
compaction barely set in. This is due to the self-tuning of two key
length scales, the thickness of the tube and the interaction range
between the centers of the discs, to be comparable to each other. When
the tube thickness is much larger than the interaction range, one
cannot avail of the attractive interaction and one obtains a highly
degenerate swollen phase. In the other extreme in which the tube
thickness is much smaller than the interaction range, one obtains a
highly degenerate compact phase -- there is a great deal of
flexibility in the relative placement of nearby tube segments. The
marginally compact phase opens up in the vicinity of the phase
transition between these two phases, when the two length scales become
comparable to each other. In the marginally compact phase, there is a
great reduction in the degeneracy of the ground state structures with
a requirement that nearby tube segments be right alongside and
parallel to each other.

Two basic requirements must be met by neighboring tube segments in the
marginally compact phase in order for them to maximally avail of the
attraction that has barely set in. First, the anisotropy of a tube
requires that neighboring tube segments be parallel to each other
rather than be perpendicular and consequently progressively separating
from each other. Second, because the range is such that the attraction
has just set in, it is crucial that neighboring segments not only be
approximately parallel to each other but right alongside each other. A
simple way of understanding how a protein is automatically poised to
be in the marginally compact phase is by noting that hydrophobicity,
which drives the self-attraction of a tube, requires that the buried
area associated with the tube be as large as possible. This drive
ensures that neighboring tube segments are placed right next to each
other to facilitate effective screening of the water.

The $\alpha$-helix is tightly packed with the main chain atoms fitting
snugly within the helix.  Likewise, in a sheet, the space between
neighboring strands is occupied by the main chain atoms. In both
cases, the scaffolding is provided by hydrogen bonds between the $N-H$
group of one amino acid and the $C=0$ group of another.  Both the tube
size and the range of the interaction are governed by the geometry of
the protein determined by quantum chemistry and more specifically the
locations of the main chain atoms. The amazingly perfect fit of the
quantum chemistry, e.g., the planarity of the peptide bond and the
lengths of the covalent and hydrogen bonds, to the structures in the
marginally compact phase is especially noteworthy.

This simple tube model is closely related to the seminal contributions
of Pauling\cite{Pauling1,Pauling2,EisenbergPNAS} and
Ramachandran\cite{Rama}. Both of them considered the protein backbone
which is the common part of all proteins. Pauling and his coworkers
explored the types of structures that are consistent with both the
backbone geometry and the formation of hydrogen bonds, which would
then provide the scaffolding for such structures. They predicted that
helices and sheets are the structures of choice in this
regard. Ramachandran and his coworkers considered the role of excluded
volume or steric interactions between nearby amino acids along the
sequence in reducing the available conformational phase space
(see~\cite{Rose} for a recent assessment of such effects on longer
sequence stretches and ~\cite{Fitzkee} for a discussion of steric
restrictions in protein folding). Astonishingly, the two significantly
populated regions of the Ramachandran plot correspond to the
$\alpha$-helix and the $\beta$-strand. Even though backbone hydrogen
bonds and steric constraints are not related to each other, they are
both promoters of helices and sheets. One might ask whether this
concurrence of events is a mere accident. The results from the simple
tube model provide a clue that the answer might be negative suggesting
that proteins, which obey physical law, may have been selected to
conform to the tube geometry through steric interactions between
nearby amino acids along the sequence and hydrogen bonds between
backbone atoms. Hydrogen bonds serve to enforce the parallelism of
nearby tube segments\cite{Cieplak}, a feature of both helices and
sheets while steric constraints emphasize the non-zero thickness of
the tube.

A more refined tube model\cite{HoangPNAS,PRE} was subsequently
introduced by incorporating the geometrical constraints of backbone
hydrogen bonds and a local bending energy penalty term. In its
simplest form, the model describes the hompolymer character of the
main backbone chain. At odds with conventional belief, it was
suggested that the gross features of the energy landscape of proteins
result from the amino acid aspecific common features of all proteins
and that protein structures lie in a marginally compact phase,
analogous to the simple tube model. This landscape is {\em
(pre)sculpted} by general considerations of geometry and symmetry and
has around a thousand broad minima corresponding to putative native
state structures. For each of these minima, the desirable funnel-like
behavior\cite{Funnel} is already achieved at the {\em homopolymer}
level. The interplay of the three energy scales, hydrophobic, hydrogen
bond, and bending energy, stabilizes marginally compact structures,
and also provides the close cooperation between energy gain and
entropy loss needed for the sculpting of a funneled energy
landscape. Further, the marginally compact phase is poised in the
vicinity of a phase transition to the swollen phase and confers
exquisite sensitivity to the structures within the phase\cite{PRE}.

In a recent manuscript, Hubner and Shakhnovich\cite{Hubner} (HS) have
presented a critique of the tube model. They state: ``The tube model
predicts that geometrical and topological factors alone, without
inclusion of more chemically detailed hydrogen bonding interactions,
determine global features of protein folds such as protein-like
secondary structure'' They then make the premise: ``Therefore, if tube
models have implications for real proteins, one would expect similar
formation, upon collapse, of helices and secondary structure motifs in
a model that accurately represented the geometric and topological
properties of amino acid chain in terms of excluded volume and
torsional degrees of freedom (as opposed to a featureless tube), but
is devoid of explicit hydrogen bonding.'' This expectation is
unfounded, since the simple tube model does predict the emergence of
secondary structure (helices and sheets) in the absence of explicit
hydrogen bonding for very short chains. While the ``compaction of a
realistic protein chain model without consideration of hydrogen
bonding does not necessarily result in helical
geometries''\cite{Hubner}, excluded volume and packing of a short tube
are sufficient to understand the emergence of protein-like secondary
structure. Furthermore, in \cite{Kamien} there was no attempt made to
explain the existence of $\beta$-sheets by invoking ``a change in the
relative sizes of the solvent and tube'', but rather the results of
the numerics were described in terms of common folding motifs.

Let us consider the coarse-graining description of HS, in which
protein coordinates representing all atoms are represented as
impenetrable hard spheres of physical radii and the degrees of freedom
associated with the water molecules are subsumed in a knowledge-based
atomic interaction potential consisting of weak non-directional Van
der Waals interactions and stronger hydrogen bonds which are highly
dependent upon geometry.  This representation of treating atoms as
hard spheres and replacing the quantum mechanics with effective
classical potentials is a coarse-graining which only works as long as
the essential ingredients underlying the system are captured
adequately. What HS demonstrate is that, in their model system,
classical potentials mimicking directional hydrogen bond formation and
Van der Waals effects promoting overall compaction lead to parts of
the sequence folding into helices. It is then not surprising that
throwing away the hydrogen bonds and retaining just the Van der Waals
interactions leads to no helix formation in the HS
model\cite{Hubner}. This result merely suggests that at this scale of
description, and for chain lengths considered by HS, the directional
hydrogen bonds play a key role.

A short self-avoiding tube subject to a self-attraction promoting
compaction, in its marginally compact phase, curls up into a helix
with a specific pitch to radius ratio\cite{MaritanNature,Kamien} close
to that observed in real protein helices and also forms zig-zag
strands which assemble into almost planar
sheets\cite{BMMT02,CPU,Thick}. Interestingly, this model, which is
sufficient for understanding individual secondary motifs of a protein,
does not require the incorporation of any classical potential
mimicking hydrogen bond formation as in the HS model. The
directionality of the hydrogen bonds is crudely captured by the
inherent anisotropy of a tube. Because the simplest description of any
chain molecule is effectively that of a tube, this result applies to
{\em any generic} polymer chain, provided it is poised in the
marginally compact part of the phase diagram. It is interesting to
note that synthetic oligomers have been shown to fold into helices
without the presence of hydrogen bonds\cite{Helix1}.

The emergence of protein-like secondary structure without the need of
explicit hydrogen bonds, for short chains within the context of the
simple tube model, does {\em not} imply, however, that we ``challenge
the view that hydrogen bonding plays an important role in protein
structure'', as stated by HS. The simple tube model, which describes a
generic polymer chain, needs to be refined in order to capture the
properties of a polypeptide chain. A more realistic yet still simple
geometrical model considers amino acid aspecific geometrical
constraints arising from the chemistry of hydrogen bonds and steric
effects and leads to assembled tertiary structures even for a chain
consisting of just one type of amino acid\cite{HoangPNAS,PRE}. It has
been shown that this refined model provides behavior in remarkable
accord with that of proteins. The marginally compact phase within this
model also provides a simple explanation for the generic formation of
amyloid\cite{Ddisease}, and elucidates the role of sequence design in
promoting the fitness of proteins in the environment of cell products
and it shows how the limited menu of geometrically determined folds
act as targets of natural selection\cite{PRE}.

Let us discuss some familiar phases of matter -- the fluid phase, the
crystal phase and the liquid crystal phase. The simplest way to
understand the fluid and crystal phases is by means of a system of
hard spheres\cite{hardsphere}. Note that the hard sphere description
in this context or, for that matter, in the HS model is itself an
emergent property\cite{Laughlin}. At low densities one obtains a fluid
phase, whereas at higher packing fractions one obtains crystalline
order. Liquid crystal phases\cite{DeGennes} arise when the objects
making up the material are no longer isotropic. Consider the formation
of smectic liquid crystals. Though Onsager showed that long enough
rods will, in general, form nematic phases independent of their
precise geometry, the same is not true for smectics.  Indeed,
sphero-cylinders undergo a nematic-to-smectic phase transition at high
enough density\cite{SK1} whereas ellipsoids do not seem to form
smectics at any density\cite{SK2}. Again, the fact that the latter
does not form the smectic phase is not indicative of the failure of
excluded volume to predict and control liquid crystalline phases;
rather, it highlights the sensitivity to the details of the specific
model, just as the HS model shows that removal of the hydrogen bonds
destroys the tendency to form helices.

Consider the sodium chloride structure adopted by ionic crystals such
as NaCl, LiCl, KBr and AgCl. The NaCl structure is a
face-centered-cubic (fcc) arrangement for the Cl ions with the sodium
ions occupying the octahedral holes. Let us consider how the structure
of the Cl ions may be determined. One can do a very careful quantum
mechanical calculation and show that this fcc structure arises from
considerations of electrovalent bonding. Alternatively, following the
pioneering work of Kepler\cite{Kepler} or the everyday experience of
grocers, one realizes that a collection of spherical cannonballs or
apples are best packed in a fcc lattice. One may then be emboldened to
suggest that considerations of packing, periodicity and the correct
symmetry (note that a packing of cubes instead of spheres would not
lead to a fcc lattice but rather a simple cubic lattice) are the
essential ingredients that determine the menu of possible crystal
structures.  In other words, the essential elements underlying the fcc
structure are not the details of the interatomic interactions or even
the quantum mechanics which describes the interactions of all matter
but rather the considerations of geometry and symmetry. It is of
course remarkable that Nature has found such a perfect fit between the
quantum interactions in NaCl and the fcc structure.

The HS exercise has a simple analogy. Let us say that a claim was made
that close packing of spheres leads to a fcc structure without
invoking charges and electrovalent bonding. Consider now doing a
calculation with effective potential energies of interaction
incorporating the electrovalent interactions on a microscopic model of
the Cl ions and finding that one recovers the fcc lattice structure
correctly. This would suggest that the model studied has enough
features to produce the right answer. Let us then imagine that on
leaving out the electrostatic interactions, one finds in this model
that the structure is no longer fcc.  Would one conclude from this
observation that the original claim that close packing of spheres
leads to a fcc structure is wrong? Of course not.  Such a result would
merely serve to show that, in the model being studied, the
electrovalent interactions were important to get the right result.
Indeed, it is well known that the structure of NaCl at the atomic
level is in fact described by electrovalent interactions. Back to the
protein context, the importance of hydrogen bonds in determining
protein structure has been recognized for more than five decades.  The
HS finding was contained in a statement in Hoang et
al.\cite{HoangPNAS}, ``Our work here underscores the importance of
hydrogen bonds in stabilizing both helices and sheets simultaneously
(without any need for adjustment of the tube thickness) allowing the
formation of tertiary arrangements of secondary motifs. Indeed, the
fine-tuning of the hydrogen bond and the hydrophobic interaction is of
paramount importance in the selection of the marginally compact region
of the phase diagram in which protein native folds are found.'' The
utility of the tube paradigm arises from its ability, in the
marginally compact phase, to capture the essential ingredients
underlying helix and sheet formation.

Consider a theoretical challenge of determining the crystal structure
for a material such as NaCl. One route would be to study the quantum
chemistry of the material in detail and calculate from first
principles that the correct structure is a face-centered-cubic
crystal.  Alternatively, one might opt to first catalog the list of
possible structures based on considerations of space-filling and
translational symmetry and then select the best fit structure from
this list. The key point is that the structure transcends the chemical
housed in it and is determined by the overarching constraints of
geometry and symmetry. The fact that many protein sequences adopt the
same fold and that the menu of possible folds is
limited\cite{Thornton} strongly suggest that similar considerations
may be at play here as well even though proteins are neither infinite
in extent nor periodic. The close packing of a flexible tube {\em in
the marginally compact phase} is then the analog of the grocer's
packing of apples for this problem.

In conclusion, we believe that the results of the HS analysis do {\em
not} disprove the tube idea.

\vspace{1cm}

\noindent {\bf Acknowledgements} We thank P. De Los Rios for
enlightening discussion. This work was supported by the 2P03B-032-25
(Ministry of Science and Informatics, Poland), PRIN MURST 2003, NASA,
NSF IGERT grant DGE-9987589 and the NSF MRSEC at Penn State.


\begin{thebibliography}{99}

\bibitem{Creighton} T.~E. Creighton, {\em Proteins: Structures and
Molecular Properties} (W.~H. Freeman and Company, New York, 1993).

\bibitem{Fershtbook} A. Fersht, {\em Structure and Mechanism in Protein
Science: A Guide to Enzyme Catalysis and Protein Folding} (W.~H. Freeman
and Company, New York, 1999).

\bibitem{Finkelbook} A.~V. Finkelstein and O. Ptistyn, {\em Protein
Physics: A Course of Lectures} (Academic Press, New York, 2002).

\bibitem{Anfinsen} C.~B. Anfinsen, Science {\bf 181}, 223 (1973).

\bibitem{Chothia1} C. Chothia, Nature {\bf 357}, 543 (1992).

\bibitem{Denton} M. Denton and C. Marshall, Nature {\bf 410}, 417 (2001).

\bibitem{Chothia2} C. Chothia, J. Gough, C. Vogel, and S.~A. Teichmann,
Science {\bf 300}, 1701 (2003).

\bibitem{Manytoone} J.~U. Bowie, J.~F. Reidhaar-Olson, W.~A. Lim, and
R.~T. Sauer, Science {\bf 247}, 1306 (1990); W.~A. Lim and R.~T. Sauer, J.
Mol. Biol. {\bf 219}, 359 (1991); D.~W. Heinz, W.~A. Baase, and B.~W.
Matthews, P. Natl. Acad. Sci. USA {\bf }89, 3751 (1992); B.~W. Matthews,
Annu. Rev. Biochem. {\bf 62}, 139 (1993).

\bibitem{Design} J.~S. Richardson and D.~C. Richardson, Trends Biochem.
Sci. {\bf 14}, 304 (1989); W.~F. DeGrado, Z.~R. Wasserman, and J.~D. Lear,
Science {\bf 243}, 622 (1989); M.~H. Hecht, J.~S. Richardson, D.~C.
Richardson, and R.~C. Ogden, Science {\bf 249}, 884 (1990); C.~P. Hill,
D.~H. Anderson, L. Wesson, W.~F. DeGrado, and D. Eisenberg, Science {\bf
249}, 343 (1990); C. Sander and R. Schneider, Proteins {\bf 9}, 56 (1991);
S. Kamtekar, J.~M. Schiffer, H.~Y. Xiong, J.~M. Babik, and M.~H. Hecht,
Science {\bf 262}, 1680 (1993); A.~P. Brunet, E.~S. Huang, M.~E. Huffine,
J.~E. Loeb, R.~J. Weltman, and M.~H. Hecht {\it et al.}\/, Nature {\bf
364}, 355 (1993); A.~R. Davidson and R.~T. Sauer, P. Natl. Acad. Sci. USA
{\bf 91}, 2146 (1994); M.~W. West, W.~X. Wang, J. Patterson, D.~J. Mancia,
J.~R. Beasley, and M.~H. Hecht, P. Natl. Acad. Sci. USA {\bf 96}, 11211
(1999); Y. Wei, S. Kim, D. Fela, J. Baum, and M.~H. Hecht, P. Natl. Acad.
Sci. USA {\bf 100}, 13270 (2003).

\bibitem{Bernal} J.~D. Bernal, Nature {\bf 143}, 663 (1939).

\bibitem{BanavarRMP} J.~R. Banavar and A. Maritan, Rev. Mod. Phys. {\bf
75}, 23 (2003).

\bibitem{HoangPNAS} T.~X. Hoang, A. Trovato, F. Seno, J.~R. Banavar, and
A. Maritan, P. Natl. Acad. Sci. USA {\bf 101}, 7960 (2004).

\bibitem{PRE} J. R. Banavar, T. X. Hoang, A. Maritan, F. Seno, A. Trovato,
Phys. Rev. E {\bf 70}, 041905 (2004).

\bibitem{Pitard} E. Pitard, T. Garel and H. Orland,
J. de Physique I {\bf 7}, 1201 (1997).

\bibitem{MaritanNature} A. Maritan, C. Micheletti, A. Trovato, and J.~R.
Banavar, Nature {\bf 406}, 287 (2000).

\bibitem{BMMT02} J.~R. Banavar, A. Maritan, C. Micheletti, and A. Trovato,
Proteins {\bf 47}, 315 (2002).

\bibitem{CPU} J.~R. Banavar, A. Flammini, D. Marenduzzo, A. Maritan, and
A. Trovato, ComPlexUs {\bf 1}, 4 (2003).

\bibitem{Thick} D. Marenduzzo, A. Flammini, A. Trovato, J.~R. Banavar,
and A. Maritan, J. Polym. Sci. Pol. Phys. {\bf 43}, 650 (2005).

\bibitem{Lezon} T. Lezon, J. R. Banavar and A. Maritan, J. Phys.
Cond. Mat. {\bf 18}, 847 (2006).

\bibitem{PL} B. Park and M. Levitt, J. Mol. Biol. {\bf 258}, 367 (1996).

\bibitem{Pauling1} L. Pauling, R.~B. Corey, and H.~R. Branson, P. Natl.
Acad. Sci. USA {\bf 37}, 205 (1951).

\bibitem{Pauling2} L. Pauling, and R.~B. Corey, P. Natl. Acad. Sci. USA
{\bf 37}, 729 (1951).

\bibitem{EisenbergPNAS} D. Eisenberg, P. Natl. Acad. Sci. USA {\bf 100},
11207 (2003).

\bibitem{Rama} G.~N. Ramachandran and V. Sasisekharan, Adv. Protein Chem.
{\bf 23}, 283 (1968).

\bibitem{Rose} R.~V. Pappu, R. Srinivasan, and G.~D. Rose,
P. Natl. Acad. Sci. USA {\bf 97}, 12565 (2000).

\bibitem{Fitzkee} N. C. Fitzkee and G. D. Rose, Protein Science {\bf
13}, 633 (2004).

\bibitem{Cieplak} J. R. Banavar, M. Cieplak and A. Maritan, Phys. Rev.
Lett. {\bf 93}, 238101 (2004).

\bibitem{Funnel} P.~E. Leopold, M. Montal, and J.~N. Onuchic, P. Natl.
Acad. Sci. USA {\bf 89}, 8271 (1992); P.~G. Wolynes, J.~N. Onuchic, and D.
Thirumalai, Science {\bf 267}, 1619 (1995); K.~A. Dill and H.~S. Chan,
Nat. Struct. Biol. {\bf 4}, 10 (1997).

\bibitem{Hubner} I.~A. Hubner and E.~I. Shakhnovich, Phys. Rev. E {\bf
72}, 022901 (2005).

\bibitem{Kamien} Y. Snir and R.~D. Kamien, Science {\bf 307}, 1067 (2005).

\bibitem{Helix1} J.~C. Nelson, J.~G. Saven, J.~S. Moore, and P.~G. Wolynes,
Science {\bf 277}, 1793 (1997); D.~J. Hill, M.~J. Mio, R.~B. Prince,
T.~S. Hughes, and J.~S. Moore, Chem. Rev. {\bf 101}, 3893 (2001); T.
Nakano and Y. Okamoto, Chem. Rev. {\bf 101}, 4013 (2001).

\bibitem{Ddisease} C.~M. Dobson, Nat. Rev. Drug Discov. {\bf 2}, 154
(2003);  D. M. Fowler et al.  PLOS Biology 4, 1 (2006).

\bibitem{hardsphere} P.~M. Chaikin and T.~C. Lubensky, {\em Principles of
condensed matter physics} (Cambridge University Press, Cambridge, 2000).

\bibitem{Laughlin} R. B. Laughlin, {\em A different universe: Reinventing
Physics from the Bottom Down} (Basic Books, New York, 2005).

\bibitem{DeGennes} P.~G. de Gennes and J. Prost, {\em The physics of
Liquid Crystals} (Oxford University Press, Oxford, 1995); S.
Chandrasekhar, {\em Liquid Crystals} (Cambridge University Press,
Cambridge, 1977).

\bibitem{SK1} A. Stroobants, H.~N.~W. Lekkerkerker, and D. Frenkel,
Phys. Rev. Lett. {\bf 57}, 1452 (1986).

\bibitem{SK2} D. Frenkel, B.~M. Mulder, and J.~P. McTague,
Phys. Rev. Lett. {\bf 52}, 287 (1984).

\bibitem{Kepler} George G. Szpiro, {\em Kepler's Conjecture: How Some of
the Greatest Minds in History Helped Solve One of the Oldest Math Problems
in the World} (John Wiley \& Sons, Inc., Hoboken, New Jersey, 2003).

\bibitem{Thornton} D.~T. Jones, W.~R. Taylor, and J.~M. Thornton, Nature
{\bf 358}, 86 (1992).

\par

\end{thebibliography}
\end{document}